A High Luminosity $e^+e^-$ Collider in the LHC tunnel to study the Higgs Boson

Alain Blondel[1], Frank Zimmermann[2]

[1]DPNC, University of Geneva, Switzerland; [2]CERN, Geneva, Switzerland

**Abstract:** We consider the possibility of a 120x120 GeV e+e- ring collider in the LHC tunnel. A luminosity of $10^{34}$/cm$^2$/s can be obtained with a luminosity life time of a few minutes. A high operation efficiency would require two machines: a low emittance collider storage ring and a separate accelerator injecting electrons and positrons into the storage ring to top up the beams every few minutes. A design inspired from the high luminosity b-factory design and from the LHeC design report is presented. Statistics of about $2 \times 10^4$ HZ events per year per experiment can be collected for a Standard Higgs Boson mass of 115-130 GeV.

## Introduction

If the Higgs boson is discovered at the LHC in the presently allowed range [1], the question may arise of performing detailed studies and precise measurements of this unique spin 0 elementary particle. The possibility to produce the Higgs boson in the s-channel to perform ultra precise studies of mass and width is one of the characteristics of the muon collider [2]. Higgs production in $e^+e^-$ collisions has been studied extensively, both for LEP and for a future high energy collider, such as TESLA [3] and the ILC [4]. The decay of Higgs bosons into pairs of known fermions will test quantitatively the hypothesis of coupling to fermion masses, while invisible final states may reveal the existence of dark matter candidates.

The maximum cross-section, and arguably the optimal centre-of-mass energy for studies of a number of Higgs boson properties, is located 10-30 GeV above the kinematic threshold for $e^+e^- \rightarrow$ ZH, i.e. $E_{cm} \sim m_H + (111 \pm 10)$ GeV [5], see Figure 1. At that energy the cross-section is of the order of 200 fb. Given that LEP was closing into the presently allowed range, and that it had been suggested to upgrade its reach up to 240 GeV $E_{cm}$ by addition of superconducting RF cavities, we have made a first investigation of the potential of a low emittance $e^+e^-$ collider in the LHC tunnel, which we call "LEP3" for convenience. The scheme presented here is largely inspired by the LHeC studies [7] of a lepton-proton collider for the LHC.

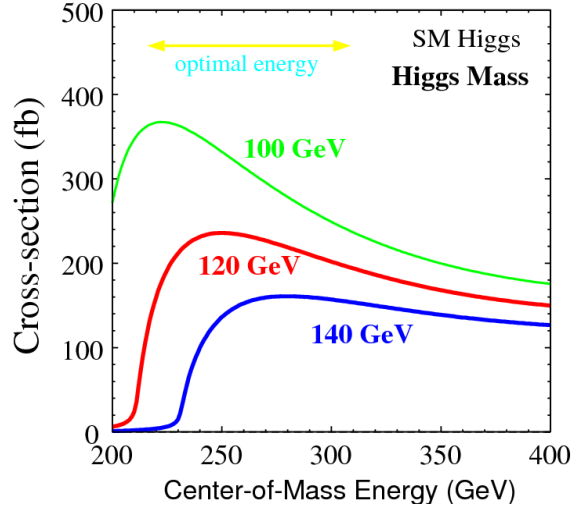

**Figure 1 Higgs boson cross section in e$^+$e$^-$ annihilation [6]**

## Ring properties

We take the LHeC design as a convenient reference to estimate the achievable peak luminosity of LEP3. The parameters of LEP, the LHeC e- ring design, and LEP3 are compared in Table 1.

Table 1: Parameters of LEP, the LHeC ring design, and LEP3 - a new electron-positron collider in the LHC tunnel, extrapolated from the LHeC design.

|  | LEP [8] [9] | LHeC ring design [7] | LEP3 |
| --- | --- | --- | --- |
| $E_b$ beam energy | 104.5 GeV | 60 GeV | 120 GeV |
| beam current | 4 mA (4 bunches) | 100 mA (2808 bunches) | 7.2 mA (3 bunches) |
| total #e- / beam | 2.3e12 | 5.6e13 | 4.0e12 |
| horizontal emittance | 48 nm | 5 nm | 20 nm |
| vertical emittance | 0.25 nm | 2.5 nm | 0.15 nm |
| $\rho_b$ dipole bending radius | 3096 m | 2620 m | 2620 m |
| partition number $J_\varepsilon$ | 1.1 | 1.5 | 1.5 |
| momentum compaction | 1.85x10$^{-4}$ | 8.1x10$^{-5}$ | 8.1x10$^{-5}$ |
| SR power / beam | 11 MW | 44 MW | 50 MW |
| $\beta_{x,y}$* | 1.5, 0.05 m | 0.18, 0.10 m | 0.15 0.0012 m |
| rms IP beam size | 270, 3.5 micron | 30, 16 micron | 55, 0.4 micron |
| hourglass loss factor | 0.98 | 0.99 | 0.65 |
| energy loss per turn | 3.408 GeV | 0.44 GeV | 6.99 GeV |
| total RF voltage | 3641 MV | 500 MV | 9000 MV |
| beam-beam tune shift (/IP) | 0.025, 0.065 | N/A | 0.126, 0.130 |
| synchrotron frequency | 1.6 kHz | 0.65 kHz | 2.98 kHz |
| average acc.field | 7.5 MV/m | 11.9 MV/m | 18 MV/m |
| effective RF length | 485 m | 42 m | 505 m |
| RF frequency | 352 MHz | 721 MHz | 1300 MHz |
| rms energy spread | 0.22% | 0.116% | 0.232% |
| rms bunch length | 1.61 cm | 0.688 cm | 0.30 cm |
| peak luminosity / IP | 1.25x10$^{32}$ cm$^{-2}$s$^{-1}$ | N/A | 1.33x10$^{34}$ cm$^{-2}$s$^{-1}$ |
| number of IPs | 4 | 1 | 2 |
| beam lifetime | 6.0 h | N/A | 12 minutes |

We assume the same arc optics as for the LHeC, which provides a horizontal emittance significantly smaller than for LEP, at equal beam energy, and whose optical structure is compatible with the present LHC machine, allowing co-existence with the LHC. Instead of the LHeC 702 MHz RF system we consider ILC-type RF cavities at a frequency of 1.3 GHz, since the latter are known to provide a high gradient and help to reduce the bunch length, thus enabling a smaller $\beta_y^*$.

A key parameter is the energy loss per turn: $E_{loss}$ [GeV] = 88.5 $10^{-6}$ $(E_b[GeV])^4/ \rho_b[m]$. The bending radius for the LHeC is smaller than for LEP, which translates into a higher energy loss than if scaled from LEP. For 120 GeV beam energy the arc dipole field is 0.1526 T. A compact magnet design as in [7] can be considered. The critical photon energy is 1.4 MeV.

We increase the ratio of RF voltage to energy loss per turn with respect to the corresponding value at LEP in order to obtain greater operational stability. We consider an RF gradient of 18 MV/m, similar to what is being assumed in the LHeC linac-ring design, and almost 2.5 times higher than the LEP gradient. An even higher gradient is technically possible, but would imply a large cryopower, the latter increasing with the square of the gradient. At 18 MV/m RF gradient the total length of the RF sections at 120 GeV beam energy is comparable to the one which had been required for LEP2 at 104.5 GeV (500 m), and the cryo power required for the collider ring is expected to be less than half the amount used for the LHC.

The unnormalized horizontal emittance is determined by the optics and varies with the square of the beam energy. We simply scale it from the 60-GeV LHeC value. The vertical emittance depends on the quality of vertical dispersion and coupling correction. We assume the vertical to horizontal emittance ratio to be similar to (slightly worse than) the one for LEP. The ultimate limit on the vertical emittance is set by the opening angle effect (emission angle of the SR photons) [10][11], which is $\varepsilon_y \geq (4.4 \times 10^{-11} T^{-1}) \beta_y B/\gamma$. With a field of 0.15 T at 120 GeV energy this limit amounts to a negligible value, below 1 fm.

The bunch length scales linearly with the momentum spread and with the momentum compaction factor and with the inverse synchrotron frequency. The bunch length of LEP3 is smaller than for LEP. Despite the higher beam energy, due to the smaller momentum compaction factor, the larger RF voltage, and the higher synchrotron frequency.

Similar to the LHeC design, the total RF wall plug power per beam is taken to be limited to 100 MW. The wall-to-beam energy conversion efficiency is assumed to be 50%. The energy loss per turn then determines the maximum beam current. At 120 GeV beam energy it is 7.2 mA or $4 \times 10^{12}$ particles per beam. Additional power will be needed for the cryoplants (a total of 10-30 MW depending on the $Q_0$ value of the cavities [7]) and for the injector/accelerator rings. The total wall plug power of the LEP3 complex would then be between 200 and 300 MW. Distributing the total charge over 3 bunches per beam each bunch contains about 1.3x1012 electrons (positrons), and the value of the beam-beam tune shift close to 0.13 is similar to the maximum beam-beam tune shift reached at KEKB. For comparison, in LEP the threshold bunch population for TMCI was about $5 \times 10^{11}$ at the injection energy of 22 GeV. For LEP3, at 120 GeV (with top up injection, see below), we gain a factor 5.5 in the threshold, which almost cancels a factor $(1.3/0.7)^3$ increase in the magnitude of the transverse wake field (of the SC RF cavities) arising from the change in wake-field strength due to the different RF frequency. We note that only about half of the transverse kick factor in LEP came from the SC RF cavities, so that the actual scaling of the threshold may be more favourable. The TMCI threshold also depends – roughly linearly - on the synchrotron tune. The LEP3 synchrotron tune is about 0.3, while in LEP at injection it was below 0.15. The higher synchrotron tune would bring a further factor of 2 in the TMCI threshold, thus raising the threshold bunch intensity to about $10^{12}$ particles. Finally, the beta functions in LEP3 at the location of the RF cavities could be designed to be smaller than those in LEP (this is already true for the beta functions in the arcs), which would further push up the instability threshold. However the bunch length at 120 GeV in LEP3 is larger than it had been at injection in LEP, so that the scaling of the TMCI threshold may be more complicated and require further investigations.

The value of 1.2 mm considered for $\beta_y^*$ could be realized by using new higher-gradient larger aperture quadrupoles based on $Nb_3Sn$ (as for HL-LHC), by a judicious choice of the free length from the IP, and possibly by a semilocal chromatic correction scheme. It is close to the value giving the maximum geometric luminosity for a bunch length of 3 mm, taking into account the hourglass effect. With a free length between the IP and the entrance face of the first quadrupole of 4 m, plus a quadrupole length of 4 m, the quadrupole field gradient should be about 17 T/m and an aperture (radius) of 5 cm would correspond to more than $20\sigma_y$.

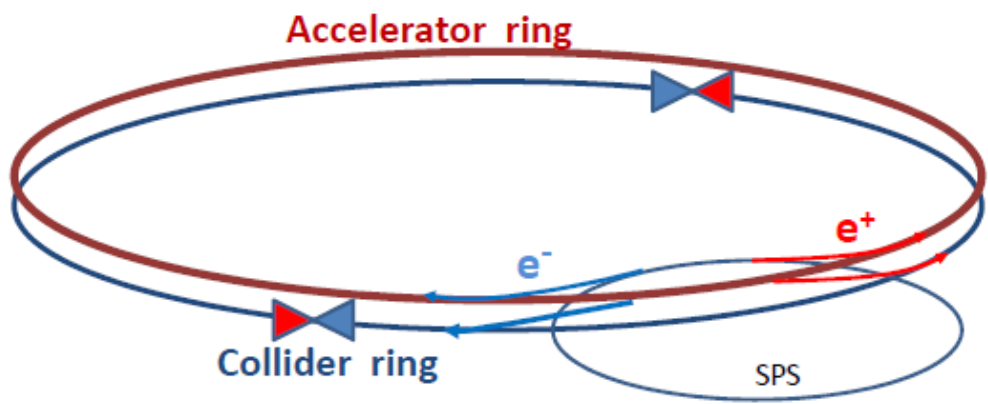

**Figure 2 Possible two ring sketch for LEP3: a first ring (accelerator ring) accelerates electrons and positrons up to operating energy (120 eV) and injects them at a few minutes interval into the low emittance collider ring in which the high luminosity $10^{34}/cm^2/s$ interaction points are situated.**

The lifetime of colliding beams at LEP was determined by radiative Bhahba scattering with a cross section of 0.215 barn [12]. It matches the directly measured lifetime [8]. At top energy in LEP2, the lifetime was dominated by the loss of particles in collisions. For a luminosity of $1.3 \times 10^{34}$ cm$^{-2}$s$^{-1}$, we find a beam lifetime $\tau_{eff}$ of 12 minutes – LEP3 would be "burning" the beams to produce physics very efficiently. In addition to the collider ring operating at constant energy, a second ring (or a recirculating linear accelerator) could be used to 'top-up' the collider continuously, as it is done in synchrotron light sources. If the top-up interval is short compared with the beam lifetime this would provide an average luminosity very close to the peak luminosity. A possible sketch is presented in Figure 2. For the top-up we need to produce about 4e12 positrons every few minutes, or of order 2e10 positrons per second. For comparison, the LEP injector complex delivered positrons at a rate of order 1e11 per second [13]. The nominal ramp rate of LEP was 500 MeV/s [14]. At the same speed acceleration to 120 GeV would take less than 4 minutes.

## Discussion

The parameter list of Table 1 allows us to draw several encouraging conclusions:

1. It is possible to envisage an electron-positron collider in the LEP/LHC tunnel with reasonable parameters operating at 120 GeV per beam with a peak luminosity of over $1 \times 10^{34}$ /cm$^2$/s in each of two interaction points while keeping the total synchrotron radiation power loss below 50 MW.
2. The luminosity lifetime is short (12 minutes). A good efficiency calls for a machine with two rings: the storage ring on one hand and an independent accelerator for the positrons and electrons that tops up the storage ring with a sufficient repetition rate to level the luminosity close to the peak value (every few minutes).
3. Operation at a luminosity of $10^{34}$ /cm$^2$/s for $10^7$ s/year leads to an integrated luminosity of 100 fb$^{-1}$/yr. For an e$^+$e$^-$ → HZ cross section of 200 fb, this yields 2 10$^4$ events per year in each experiment (we have assumed 2), allowing precise measurements of the Higgs Boson mass, cross-section and decay modes, even invisible ones. It would also provide more than $10^6$ WW events per year in each IP. This machine would have similar or better performance than a linear collider operating at the same energy, and would reach it more economically.

## Outlook

This first look at LEP3 is encouraging and should motivate a dedicated investigation.

Among the many questions that should be addressed in more detail: 1) a comparison of cost and performance for the proposed double ring separating the accelerator and collider and for a single combined ring; 2) a total of about 15 GV of RF acceleration is needed : 9 GV for the storage ring and 6 GV for the accelerator - it will be necessary to determine the optimum RF gradient as a compromise between cryopower and space requirement, and the optimum RF frequency with regard to impedance, RF efficiency and bunch length [in this paper we consider the use of high-frequency ILC-type cavities]; 3) the LHeC lattice has reduced the effective bending radius compared with LEP while one would rather like to increase it instead; 4) the performance may perhaps be further improved by using even smaller value of $\beta^*_y$ and e.g. the technique of crab waste-crossing[15]; 5) on the physics side one should review what measurements on the Higgs boson can be made at the LHC and with which precision, and whether LEP3 is really the complementary machine one needs; 6) the performance at 91.2 $E_{cm}$ (the Z peak), possibly with polarized beams, should be looked at; 7) the co-habitation of such a double machine with the LHC would require careful examination of the layout of both machines - for the single LHeC ring no show-stopper has been found [7]; 8) the suitability of the existing LHC detectors (or the desirability of new ones) for this physics and for the additional equipment needed (low beta insertions and luminosity monitors) should be investigated; 9) the ramping speed of the accelerator ring; 10) the positron source; 11) the limit on the single bunch charge; 12) the top-up scheme, e.g. injecting new bunches at full intensity or refilling those already colliding; and 13) the alternative possibility of building a new larger tunnel and storage ring(s) with twice the LEP/LHC circumference, which we call DLEP. Possible DLEP parameters are listed in Table 2, alongside those for LEP3. Naturally, in the long-distant future a DLEP tunnel could also house a proton collider ring with a beam energy about four times higher than the LHC (assuming two times stronger magnets). Rings with a circumference of 30.6 km and 50 km were proposed during the LEP design in 1979 and 1976, respectively, requiring part of the tunnel to be located in the rocks of the Jura, at a depth of 860 m under the crest [16]. The circumference of DLEP should be optimized in order to avoid Hirata-Keil resonances [17] in case the machine is later used for lepton-hadron collisions with the LHC.

Table 2: Comparison of LEP3 parameters with those for a new ring of twice the circumference: DLEP.

|  | LEP3 | DLEP |
|---|---|---|
| $E_b$ beam energy | 120 GeV | 120 GeV |
| circumference | 26700 m | 53400 m |
| beam current | 7.2 mA (3 bunches) | 14.4 mA (60 bunches) |
| total #e- / beam | 4.0e12 | 16.0e12 |
| horizontal emittance | 20 nm | 5 nm |
| vertical emittance | 0.2 nm | 0.05 nm |
| $\rho_b$ dipole bending radius | 2620 m | 5240 m |
| partition number $J_\epsilon$ | 1.5 | 1.5 |
| momentum compaction | $8.1 \times 10^{-5}$ | $2 \times 10^{-5}$ |
| SR power / beam | 50 MW | 50 MW |
| $\beta_{x,y}^*$ | 0.15, 0.0012 m | 0.2, 0.002 m |
| rms IP beam size | 55, 0.4 micron | 32, 0.3 micron |
| hourglass loss factor | 0.65 | 0.91 |
| energy loss per turn | 6.99 GeV | 3.5 GeV |
| total RF voltage | 9000 MV | 4600 MV |
| beam-beam tune shift (/IP) | 0.126, 0.130 | 0.10, 0.10 |
| synchrotron frequency | 2.98 kHz | 0.8 kHz |
| average acc.field | 18 MV/m | 11 MV/m |
| effective RF length | 505 m | 418 m |
| RF frequency | 1300 MHz | 1300 MHz |
| rms energy spread | 0.232% | 0.164% |
| rms bunch length | 0.30 cm | 0.15 cm |
| peak luminosity / IP | $1.33 \times 10^{34}$ cm$^{-2}$s$^{-1}$ | $1.63 \times 10^{34}$ cm$^{-2}$s$^{-1}$ |
| number of IPs | 2 | 2 |
| beam lifetime | 12 minutes | 19 minutes |

## Acknowledgements

We would like to thank Max Klein for an essential invitation to coffee. A.B. is grateful to Don Summers for a great (but dry) discussion at a Mexican restaurant in Oxford (Mississippi), and to many colleagues for their encouragements. F.Z. would like to thank Katsunobu Oide for a careful reading of the manuscript and for raising several important questions. We also thank Mike Koratzinos for helpful comments and corrections.